\documentclass[12pt,a4paper]{article}
\pdfoutput=1

\usepackage{graphicx}
\usepackage{bm}
\usepackage{paralist}  

\newcommand{\latexOrPdflatex}[2]{\ifx\undefined\pdfoutput%
#1%
\else%
#2%
\fi}

\latexOrPdflatex{
\newcommand{\href}[2]{#2}
}{
\usepackage[colorlinks,pagebackref,citecolor=blue,bookmarks=true,pdfstartview=FitH,pdfstartpage=1,pdftitle={Self-config from ML-Perspective},pdfauthor={Wolfgang Konen}]{hyperref}		
}

\newcommand{\eqref}[1]{Eq.~(\ref{#1})}

\newenvironment{zitat}{\begin{quote}\begin{em}\begin{small}}{\end{small}\end{em}\end{quote}}

\title{%
\vspace{1.cm} 
Self-configuration from a Machine-Learning Perspective
}
\author{Wolfgang Konen \\ \\
 Institute for Informatics, Cologne University of Applied Sciences \\
 Steinm\"ullerallee 1, D-51643 Gummersbach, Germany \\
 \texttt{\href{http://www.gm.fh-koeln.de/~konen}{http://www.gm.fh-koeln.de/{\footnotesize$\sim$}konen}} \\
 \texttt{\href{mailto:wolfgang.konen@fh-koeln.de}{wolfgang.konen@fh-koeln.de}}
}

\date{}

\begin{document} 

\maketitle 

\begin{center}
	Contribution to {\bf Dagstuhl Seminar 11181}\\ 
	{\bf "Organic Computing - Design of Self-Organizing Systems"}\\
	organized by \\
	Kirstie Bellmann, Andreas Herkersdorf and Michael G. Hinchey\\[1.2cm]
\end{center}

\begin{abstract}
The goal of machine learning is to provide solutions which are trained by data or by experience coming from the environment. Many training algorithms exist and some brilliant successes were achieved. But even in structured environments for machine learning (e.g. data mining or board games), most applications beyond the level of toy problems need careful hand-tuning or human ingenuity (i.e. detection of interesting patterns) or both. We discuss several aspects how self-configuration  can help to alleviate these problems.  
One aspect is the self-configuration by tuning of algorithms, where recent advances have been made in the area of SPO (Sequential Parameter Optimization). Another aspect is the self-configuration by pattern detection or feature construction. Forming multiple features (e.g. random boolean functions) and using algorithms (e.g. random forests) which easily digest many features can largely increase learning speed. However, a full-fledged theory of feature construction is not yet available and forms a current barrier in machine learning.  
We discuss several ideas for systematic inclusion of feature construction. This may lead to partly self-configuring machine learning solutions which show robustness, flexibility, and fast learning in potentially changing environments.  
\end{abstract}


\section{Introduction} \label{sec:introduction}

How do we learn in new environments? It is a striking feature of human behaviour that humans can adapt rapidly to new situations or environments, can spot important patterns if only few examples are given and can perform meaningful generalizations. Computing machines show a much poorer performance in these disciplines. This is not only true for complex environments with noisy sensory information, but also for  very clean and structured data. 

As examples we will study in this work two cases with fairly structured information: board games and data mining.

Games are an ideal testbed for the study of learning and self-configuring systems. It is not so much that we are interested in construction world-leading AI players for complex games like chess (this problem is mainly solved today). The interesing point is nicely formulated by Simon Lucas \cite{Luca08}:
\begin{zitat}
The immediate goal of the research ... is to study the effectiveness of machine learning approaches
to game playing: how well a machine can learn to play, rather than how well we can program it to play. In particular we are interested in how well the system can learn to play without any expert tuition, and without recourse to an expert opponent to practice against.
\end{zitat}

In games you need to discover a strategy. A strategy can be the detection of and reaction on the right pattern / feature. In principle, neural networks and reinforcement learning (RL) could solve the problem because they  can learn arbitrary features. Indeed, some remarkable successes were achieved, e.g. Tesauro's TD-Gammon \cite{Tesa94a}. Pollack and Blair showed an alternative approach based co-evolution \cite{PoBl98}, which was shortly followed by many neuroevolution contributions, starting with the work of Chellapilla and Fogel \cite{CheF99}. 

But there are other cases in RL, where the learning progress is disappointingly slow \cite{Tesa92,Kone09a}, even when only simple games are considered.
We have shown in \cite{Kone09a} examples from reinforcement learning where a simple toy problem is not solvable with the wrong features, but quickly solvable with the right features. 
It would be desirable to have a mechanism with which one can learn or construct such interesting features. We show below a general approach based on N-tuple systems, which was recently applied by Lucas \cite{Luca08} to board game problems with remarkable success.



In data mining, our second example area, we have for small problems (e.g. UCI data) or for special problems robust learning solutions  (e.g. random forests). 
But no solution is out there which works equally robust on a large variety of problems or on larger problems.
For larger problems we usually still need careful hand-tuning or ingenuity (detect patterns) or both.
It is the aim to replace some of this by self-configuration. 
Advances in this domain will have numerous applications, especially in the area of online data mining (stream data mining) with its often non-stationary environment conditions.

Recent data-mining developments try to use larger amount of features (many parallel features) and bring the optimal use of features into the optimization loop, as we try in our current research project SOMA (Systematic Optimization of Models in IT and Automation).

Other developments, e.g. the IBM Watson project, where a supercomputer learns to answer questions from the Jeopardy quiz \cite{Watson10}, rely on massively parallel models  with multi-hypothesis testing and multiple confidence estimation 

In this paper we try to advocate the following (still unproven) conjecture: 
\begin{zitat}
Finding good features and using abundant features is key for learning in high-dimensional input spaces.
\end{zitat}

The rest of the paper is organized as follows: Sec.~\ref{sec:games} looks at self-configuration and self-learning (w/o teacher) in games and looks specifically at the N-tuple approach.
Sec.~\ref{sec:DM} is about self-configuration in data mining (DM), it discusses the desirable characteristcs of self-configuring, robust, and flexible DM solutions. It looks briefly at tuning and at feature construction in DM. Finally in Sec.~\ref{sec:discuss} we discuss our findings and present a set of open research questions.

\section{Self-configuration in games} \label{sec:games}
	\subsection{Strategy example}
Reinforcement learning is a remarkable example of self-organization in game play: Just by playing the game against itself, an agent can learn emergent behaviour, e.g. to play on a master level for some games, as was shown with Tesauro's  TD-gammon \cite{Tesa94a}. But reinforcement learning alone is not the complete solution. It is necessary to detect / construct the right features, otherwise the success may be slowed down dramatically or may be completely blocked \cite{Kone09a}.  

A simple example might illustrate the point: In the game Nim-3 either player can take 1, 2 or 3 pieces from a collection of initially $n$ pieces and winner is the one who picks the last piece. The right feature to detect is of course: "Leave your opponent $m$ stones, where $m$ is a number divisible by 4." As long as you do not spot this pattern or feature, only complicated calculations find the right move for sufficiently large $n$. With this feature, the formulation of a winning strategy is trivial for arbitrary $n$.

\begin{figure}[ht]
	\centering
	\includegraphics[width=0.5\textwidth]{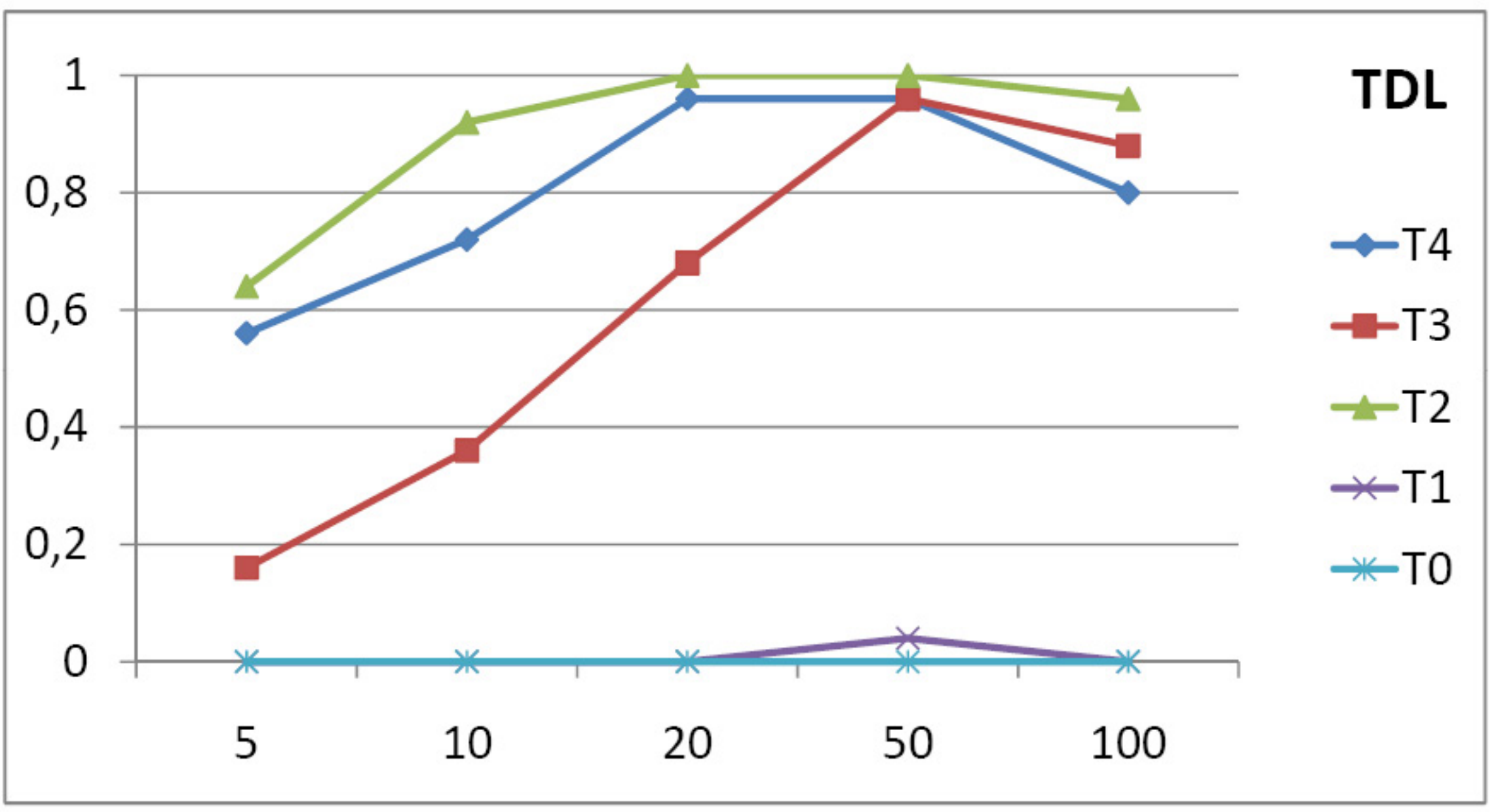}
	\caption{The influence of different feature sets T0,...,T4 on the speed of TD-learning. The x-axis shows the number of trainng games (in thousands), the y-axis shows the success rate of the trained agent \cite{Kone09a}.		\label{fig:Feature-TDL}	}
\end{figure}

We investigated in \cite{Kone09a}  a similar but non-trivial case for a small game (TicTacToe) and found similar results (Fig.~\ref{fig:Feature-TDL}): Some feature sets (T2, T4) make TD-learning very fast, while others (T0, T1) completely block successful TD-learning, among them the raw board position (T0). The feature sets T2 and T4 contained human-designed features. 

\begin{figure}[ht]
	\centering
	\includegraphics[width=0.8\textwidth]{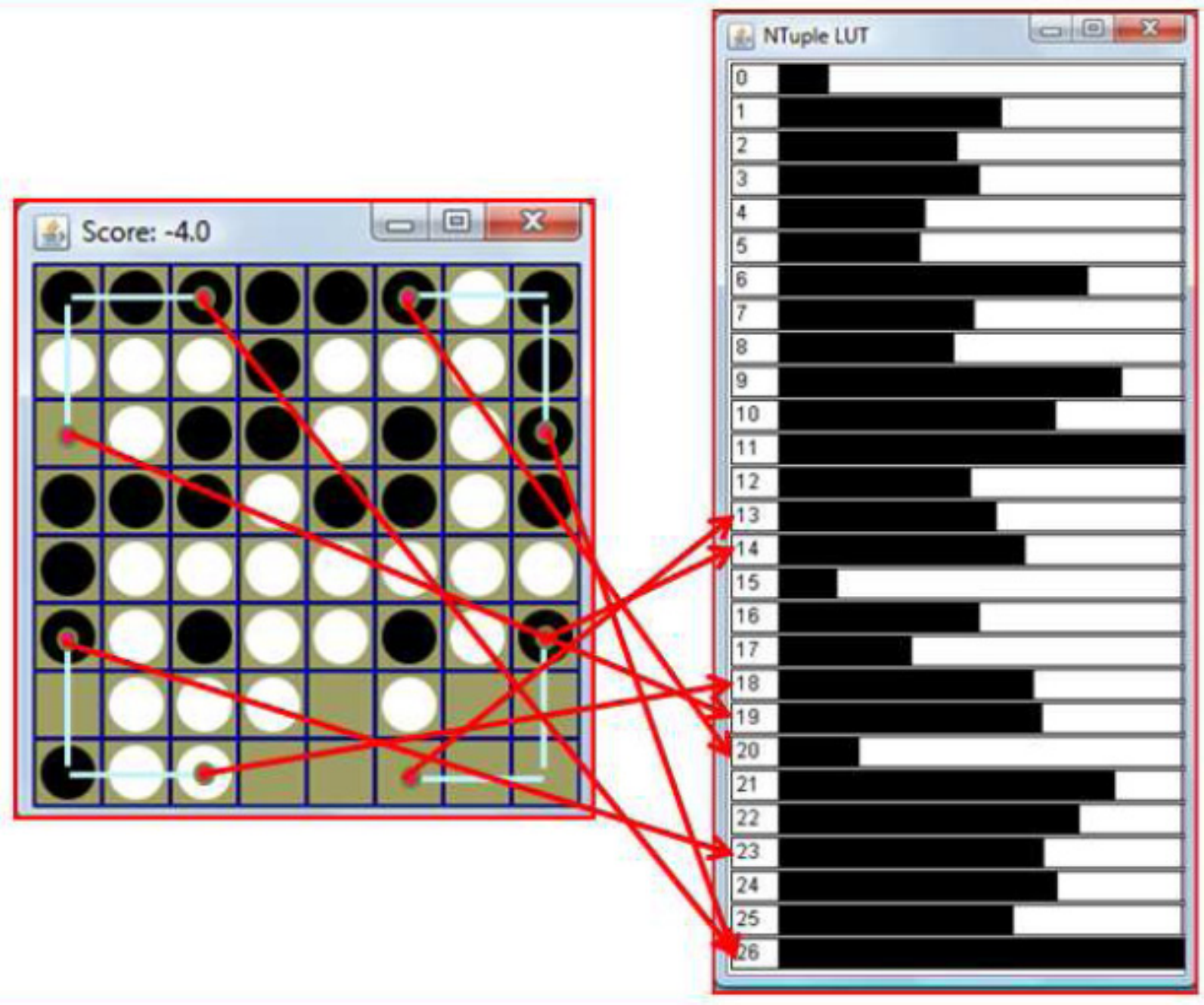}	
	\caption{3-tuple example from \cite{Luca08}. LUT cells are indexed (adressed) with the ternary code \{white=0, empty=1, black=2\}. Each LUT cell $d$ has a weight $l(d)$ assigned (black bar; the higher, the better for Black).
	\label{fig:Ntuple} }
\end{figure}

	\subsection{N-tuple systems}
The question is how to find such or other features given a new task or a new game \textbf{without} the need for human design. Lucas \cite{Luca08} has recently proposed to use the old idea of N-tuple systems for the first time also for the detection / construction of useful features in games.  
As an example he considered the game Othello: An N-tuple is here a chain of length $N$ formed by a random walk on the board. Fig.~\ref{fig:Ntuple} shows a 3-tuple example. The piece situation along the chain is taken as adress into a LUT (look-up table). If the game possesses symmetries (here: 8-fold, reflection and rotation) then all symmetric N-tuple positions do also activate a LUT entry. Therefore we see 8 red arrows in Fig.~\ref{fig:Ntuple}. Each LUT entry $d$ has a weight $l(d)$ connected with it. Given a board position $b$ which activates the LUT entries in set $D(b)$, the output of the N-tuple is simply the sum
\begin{equation}
	v(b) = \sum_{d\in D(b)}{l(d)}
	\label{eq:ntupleSum}
\end{equation}
Given $K$ such randomly formed N-tuples, the output of the network given a board position $b$ is the sum over all N-tuples
\begin{equation}
	V(b) = \sum_{k=1}^{K}{v_k(b)}
	\label{eq:networkSum}
\end{equation}
The goal is that $V(b)$ approximates the game's value function, i.e. that it assigns to each board position $b$ the correct probability of Black to win from this $b$. The weights $l(d)$ are trained by temporal difference learning (TD-learning). 

The important message from \cite{Luca08}: Lucas formed randomly 30 N-tuples and he succeeded in generating a strong-playing Othello agent within only 1250 games (!) of training. This is remarkable, since other reinforcement learning approaches to Othello usually need millions of games: van Eck  and van Wezel \cite{vEck05} use Q-learning and need 15,000,000 games and Szubert et al. \cite{Szub09} use coevolution together with TD-learning and need 4,500,000 games.

It seems that the use of abundant features greatly influences the speed of learning, the speed at which self-organization of emergent behaviour appears. Why is this the case? At present we can only formulate hypotheses: As Lucas \cite{Luca08} notes, the N-tuple system 
\begin{zitat}
... is somewhat similar to the kernel trick used in support vector machines (SVMs) and is also related to Kanerva's sparse distributed memory model \cite{Kane88}. The low dimensional board is projected into a high dimensional sample space by the N-tuple indexing process.  
\end{zitat}
%
%
The high-dimensional sample space can help to make the game strategy easier to learn. It reduces the probability that weights receive conflicting signals from different input situations. The indexing operation into the LUT performs a non-linear mapping to high-dimensional feature space. The weights of the LUT can be seen as the weights of a single-layer perceptron.
	  			
In fact, the N-tuple system has an architecture close to Rosenblatt's perceptron \cite{Rose62}. The original perceptron had hidden units which formed fixed random boolean functions on the subset of the input. N-tuples are also random boolean functions on a subset of the input. This means that N-tuple systems share the strengths (fast training) and the weaknesses (theoretical limitations, "parity problem") with the perceptron. But note that the N-tuple system has learned a rather complex Othello value function. The main differences between the original perceptron and Lucas' approach are that the target signal is delivered by the TD-learning concept and that the number of hidden units (number of LUT entries in all LUTs) is potentially very large. 
 Is it that the vast amount of LUT entries has an influence on the self-organizational capabilities of the system? We want to investigate this question in the near future. 


\section{Self-configuration in machine learning and data mining} \label{sec:DM}
	\subsection{What is desirable?}
	Machine learning models have made considerable progress during the last decades: Random forests (RF) are more robust and flexible than single decision trees; support vector machines (SVM) can better handle large numbers of input variables and build implicitly high-dimensional feature spaces. But still: there is no free lunch, expert knowledge is required to get high quality results for complicated, perhaps noisy tasks. Usually the expert is needed to (a) build the right features (preprocessing), (b) select the right features, and (c) select the right model and tune model and preprocessing parameters.  
	
	It would be desirable to have a complete solution which is able to configure itself on the task just by looking at the data and by self-configuring the options in (a) - (c). This becomes even more important if we want to build systems for  {\em online data mining} or {\em stream data mining}, where new data are continuously streaming in and the underlying process might also be non-stationary, thus requiring self-configuration or re-calibration in daily routine.
	
	\subsection{Tuning}
Tuning, i.e. parameter adjustment, can be seen as a form of self-configuration. Although the problem is simply stated: "There are $N$ adjustable parameters - find the best values for them given the current task", it is seldom done systematically in its general form. Why? - It quickly gets complex, if $N$ becomes larger (curse of dimensionality) and/or if it contains mixed boolean, integer and real-valued parameters. Furthermore, for complex tasks the modeling step is often time-consuming. Only few parameter adjusting runs may be feasible within the given budget. 

\begin{figure}[htbp]
	\centering
	\includegraphics[width=1.0\textwidth]{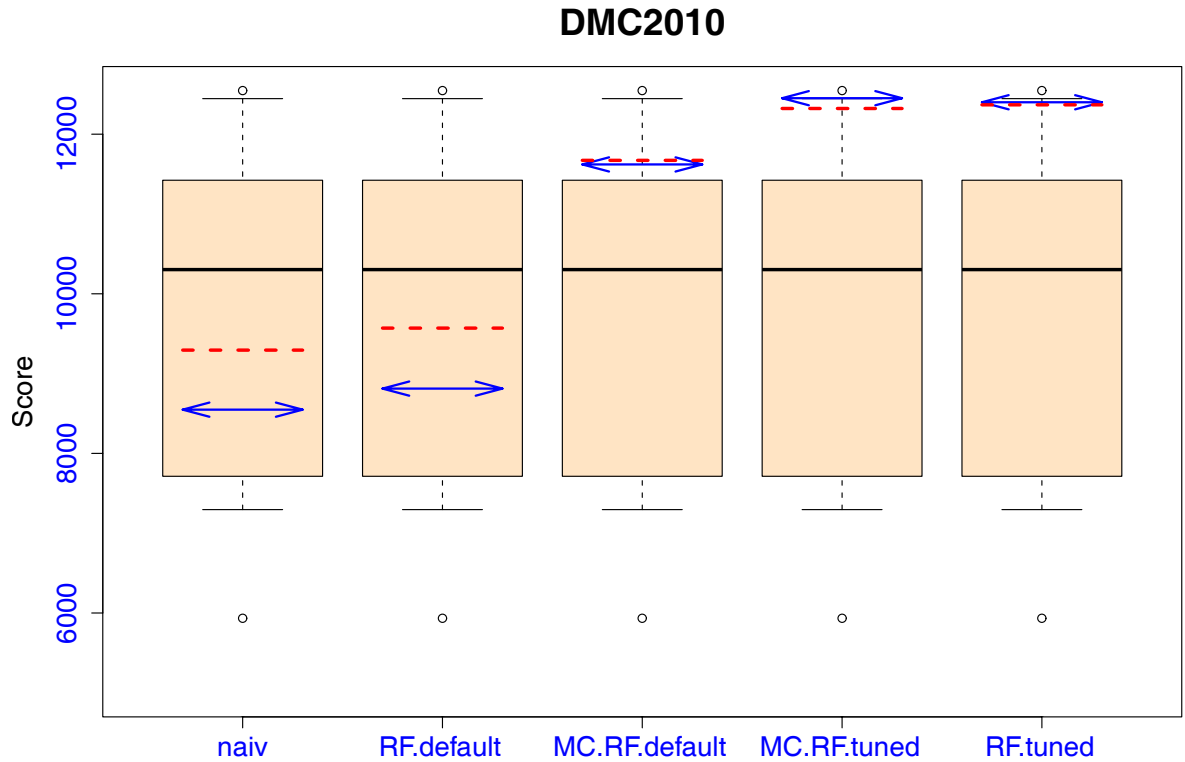}
	\vspace{-0.5cm}
	\caption{Our tuning results for the  Data Mining Cup 2010 benchmark \cite{Kone11b}. The red dashed lines
show the score of our models on the training data (10-fold cross validation),
the blue arrows show the score on independent test data. The boxplot shows for comparision the spread
of score among the competition participants. It is clearly visible, that the models with default parameter settings (columns 2 and 3) do not produce results of high quality, while the tuned models (columns 4 and 5) have high-quality results close to the winner of the Data Mining Cup 2010 benchmark (MC = MetaCost, RF = Random Forest).}
	\label{fig:DMC2010-tuning}
\end{figure}

Recently, some advances in tuning have been made: Bartz-Beielstein's sequential parameter optimization (SPO) \cite{Bart09f,Bart10e} builds meta models (surrogate models) with the help of Kriging or other algorithms 
and makes it possible to get good optimization results within only a few runs of the real model. Other interesting tuning methods are CMA-ES, REVAC, BFGS, and other. A recent comparison of different tuners for DM can be found in \cite{Kone11b}. We present examplary results from \cite{Kone11b} in Fig.~\ref{fig:DMC2010-tuning}.

	\subsection{Feature construction and feature selection}
	According to Fayyad's definition \cite{Fayy96}: 
\begin{zitat}
	Data mining is the nontrivial process of identifying valid, novel, potentially useful, and ultimately understandable patterns in data. 
\end{zitat}	
	we know that data mining is tightly connected with pattern finding.

There are numerous (infinitely many) ways to construct features from input data	
Besides the transformation for (spatial or temporal) continuous signals (Fourier transform, wavelet transforms, other), which we do not consider here, some examples are
\begin{enumerate}
	\item N-tupel systems (see above)
	\item GP (genetic programming) \cite{Koza92} 
	\item PCA, GHA (variance, assumes linearity)  \cite{Oja82,Sanger89}
	\item SFA	(slowness, assumes linearity in a higher dim space) \cite{Wis98a,WisSej2002}
	\item Kernel PCA \cite{Scho98}, KHA (nonlinear) \cite{KimF03}
\end{enumerate}
Example 1 and 2 are of the class "form many features 'at random' and find somehow the ones which are good for the task", while the remaining examples belong to the class "form features guided by some general principles and select the important ones".
	
\begin{figure}[htbp]
	\centering
	\includegraphics[width=0.75\textwidth]{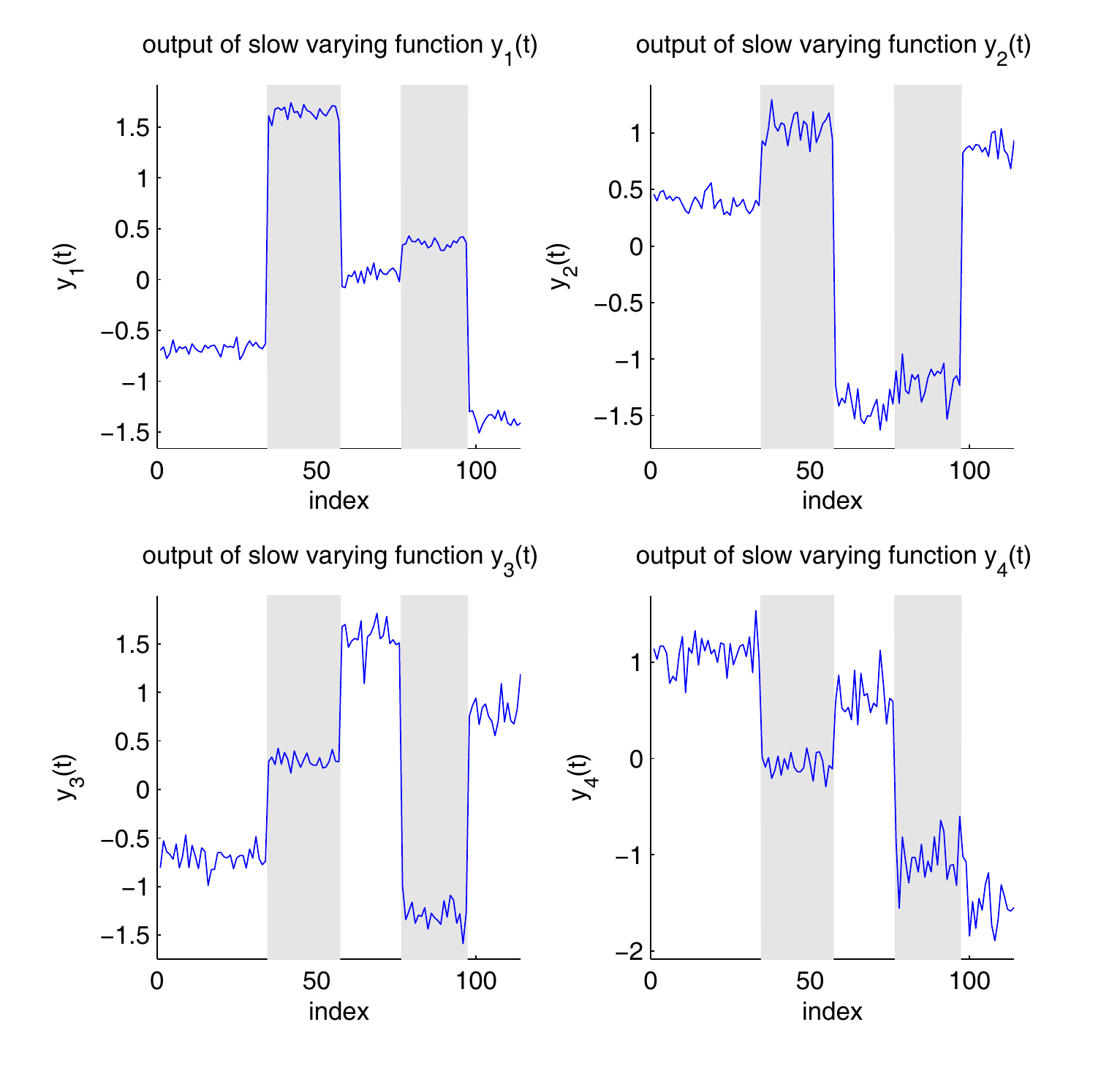}
	\vspace{-0.5cm}
	\caption{Gesture classification with SFA: Output of the first four SFA feature detectors $y_1, y_2, y_3, y_4$ for the different gesture classes (ordered along the x-axis) \cite{Koch10a}.} 
	\label{fig:gesture-SFA}
\end{figure}

\begin{figure}[htbp]
	\centering
	\includegraphics[width=0.6\textwidth]{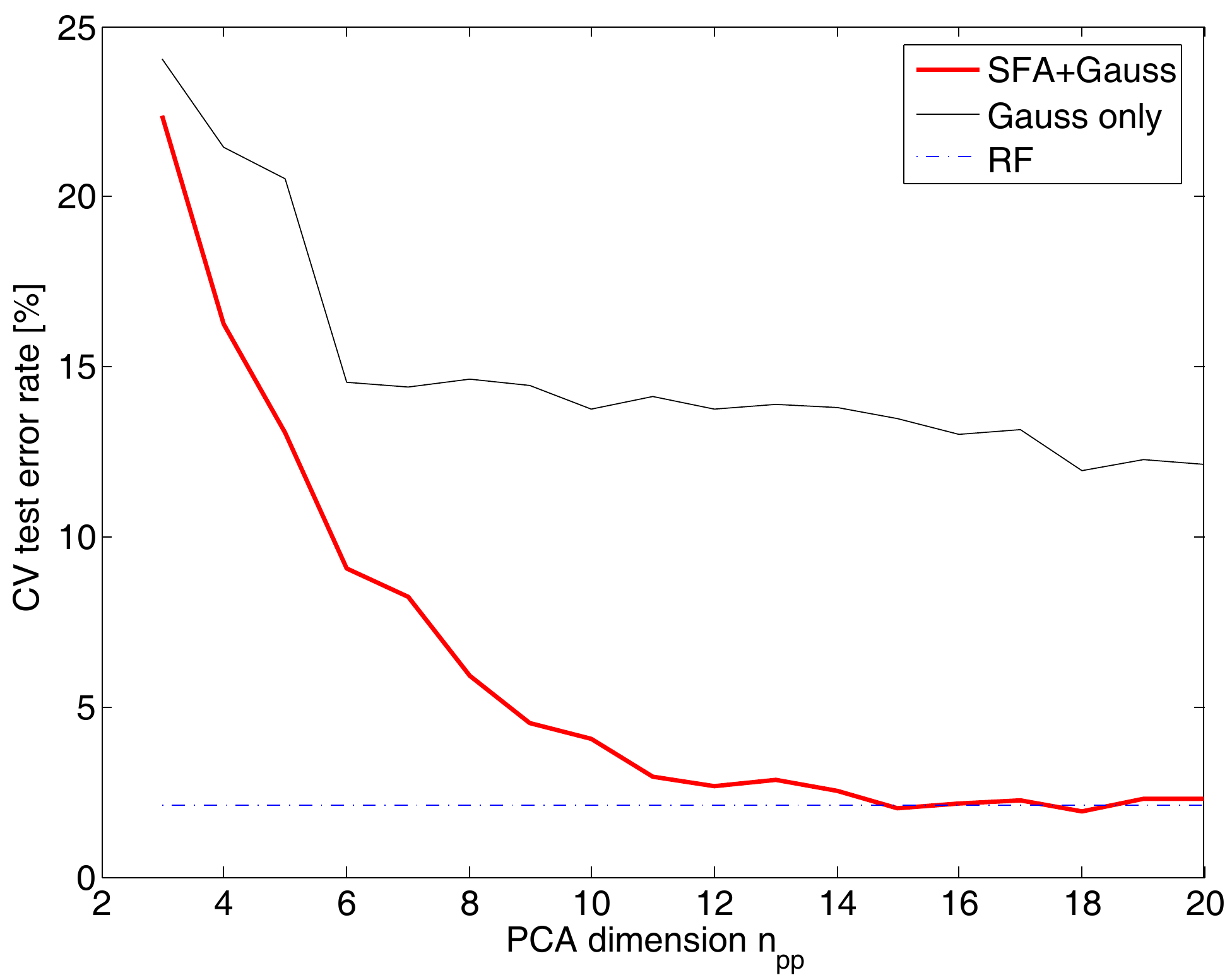}
	\vspace{-0.5cm}
	\caption{"SFA+Gauss": Classification error in gesture recognition with SFA working on $n_{pp}$ preprocessed PCA features. "Gauss only": Gauss classifier on the PCA features alone. "RF": random forest classifier. }
	\label{fig:gesture-error}
\end{figure}

We had some good experience with SFA on a gesture classification task which could not be successfully solved by a solely PCA-based feature construction, see Fig.~\ref{fig:gesture-SFA} and Fig.~\ref{fig:gesture-error} \cite{Koch10a}.	

\section{Discussion} \label{sec:discuss}
Some evidence has been presented that feature construction can be very relevant for success and fast convergence (both in time and in number of training examples) of machine learning problems. We have no proof that this is always the case, but it is my working hypothesis that: 
\begin{quote}
"Self-configuring machine learning systems will require one or several modules for feature construction if they are to work robustly and flexible on a large variety of problems."   
\end{quote}
In order to strengthen this hypothesis, one should build machine learning systems containing explicit feature construction modules and compare their performance with systems where the construction of useful representations (features) is implicit and is delegated to the model-building step.

However, there is currently no generally accepted theory or framework of feature construction for arbitrary data mining or machine learning problems. Even worse, the problem is ill-defined right at the beginning, because there is no way to define exhaustively what a feature might look like given a set of $N$ input variables. 
But the problem seems important and although I do not have a full solution right now, I would like to discuss it more deeply here in the workshop and perhaps we can bring together a framework of interesting ideas for self-configuration in the context of feature construction. To start the discussion I have put together some questions:
\begin{itemize}
	\item Which road to follow: random feature formation (try many and throw away many) or more careful feature construction driven by (complex) guiding principles?
	\item Is GP (Genetic Programming) a solution?
	\item Which general guiding principles can be used for feature formation / construction?
	\begin{compactitem}
	\item variance (PCA)
	\item slowness (SFA)
	\item information gain or other, guided by supervised information
	\item ...
	\end{compactitem}
	\item Self-configuration: Given a large number of constructed features (from different approaches), are there {\em fast} procedures to decide which features are most promising for the given task? 
	\item Are there procedures which might switch features on-line if the task is non-stationary?
	\item Which is the role of hierarchy? Could we initially use random feature formation, gain some experience from the environment and then combine the most promising features to form more complex features which are better adapted to the task? 
	\item Can one translate the random N-tuple approach to real-valued variables?
\end{itemize}

\section{Conclusion}

Self-configuration in machine learning is still difficult, if problem scale gets bigger.
Tuning should be part of the solution, but it is not the solution.
Feature construction seems not solvable in the general sense, but it should not be neglected.
Conjecture (to be proven): 
\begin{zitat}
Rich-enough feature sets lead to robustness, flexibility and fast learning in potentially changing environments.
\end{zitat}
Advances in this area will have numerous applications, especially in the area of online data mining (stream data mining) with often non-stationary environment conditions.

\section{Acknowledgements}
\begin{minipage}{0.75\textwidth}
This work has been supported by the Bundesministerium f\"ur Bildung und Forschung (BMBF) under the
grant SOMA (AiF FKZ 17N1009, "Ingenieurnachwuchs")
and by the Cologne University of Applied Sciences under the research focus grant COSA. 
\end{minipage}
\begin{minipage}{0.25\textwidth}
\begin{flushright}
	\includegraphics[width=\textwidth]{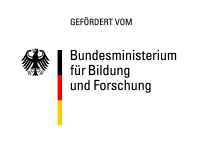}
\end{flushright}
\end{minipage}


\begin{thebibliography}{KKF{\etalchar{+}}11}

\bibitem[BB10]{Bart10e}
Thomas Bartz-Beielstein.
\newblock \href{http://arxiv.org/abs/1006.4645}{{SPOT}: An {R} Package For
  Automatic and Interactive Tuning of Optimization Algorithms by Sequential
  Parameter Optimization}.
\newblock Technical Report arXiv:1006.4645. CIOP Technical Report 05-10,
  Cologne University of Applied Sciences, June 2010.

\bibitem[BBLP10]{Bart09f}
Thomas Bartz-Beielstein, Christian Lasarczyk, and Mike Preuss.
\newblock The sequential parameter optimization toolbox.
\newblock In Bartz-Beielstein et~al., editors, {\em Experimental Methods for
  the Analysis of Optimization Algorithms}, pages 337--360. Springer, Berlin,
  Heidelberg, New York, 2010.

\bibitem[CDF99]{CheF99}
K.~Chellapilla and D.~D.~Fogel.
\newblock Evolving neural networks to play checkers without expert knowledge.
\newblock {\em IEEE Transactions on Neural Networks}, pages 1382--1391, 1999.

\bibitem[FB{\etalchar{+}}10]{Watson10}
D.~Ferrucci, E.~Brown, et~al.
\newblock
  \href{http://www.stanford.edu/class/cs124/AIMagzine-DeepQA.pdf}{Building
  {W}atson: An Overview of the {DeepQA} Project}.
\newblock {\em {AI} Magazine}, page~60, 2010.

\bibitem[FPSS96]{Fayy96}
U.~Fayyad, G.~Piatetsky-Shapiro, and P.~Smyth.
\newblock From data mining to knowledge discovery in databases.
\newblock {\em AI Magazine}, 17:37--54, 1996.

\bibitem[Kan88]{Kane88}
P.~Kanerva.
\newblock {\em Sparse Distributed Memory}.
\newblock MIT Press, Cambridge, Mass., 1988.

\bibitem[KBB09]{Kone09a}
Wolfgang Konen and Thomas Bartz-Beielstein.
\newblock \href{http://doi.acm.org/10.1145/1570256.1570375} {Reinforcement
  learning for games: failures and successes}.
\newblock In {\em Proc. GECCO'09, Montreal}, pages 2641--2648, New York, NY,
  USA, 2009. ACM.

\bibitem[KFS03]{KimF03}
Kwang~In Kim, Matthias~O. Franz, and Bernhard Sch\"{o}lkopf.
\newblock Kernel {H}ebbian algorithm for iterative kernel principal component
  analysis.
\newblock Technical report, Max Planck Institute for Biological Cybernetics,
  2003.

\bibitem[KKF{\etalchar{+}}11]{Kone11b}
Wolfgang Konen, Patrick Koch, Oliver Flasch, Thomas Bartz-Beielstein, Martina
  Friese, and Boris Naujoks.
\newblock
  \href{http://maanvs03.gm.fh-koeln.de/webpub/CIOPReports.d/Kone11b.d/tdm-cioprep.pdf}{Tuned
  Data Mining: A Benchmark Study on Different Tuners}.
\newblock CIOP Technical Report 02-11, Cologne University of Applied Sciences,
  Feb 2011.

\bibitem[KKH10]{Koch10a}
P.~Koch, W.~Konen, and K.~Hein.
\newblock
  \href{http://www.gm.fh-koeln.de/~konen/Publikationen/WCCI-10_wii-gesture.pdf}{Gesture
  Recognition on Few Training Data using Slow Feature Analysis and Parametric
  Bootstrap}.
\newblock In {\em 2010 International Joint Conference on Neural Networks},
  2010.

\bibitem[Koz92]{Koza92}
J.R. Koza.
\newblock {\em Genetic Programming: On the Programming of Computers by Means of
  Natural Selection}.
\newblock MIT Press, Cambridge MA, 1992.

\bibitem[Luc08]{Luca08}
Simon~M. Lucas.
\newblock Learning to play othello with n-tuple systems.
\newblock {\em Australian Journal of Intelligent Information Processing}, pages
  1--20, 2008.

\bibitem[Oja82]{Oja82}
E.~Oja.
\newblock A simplified neuron model as a principal component analyzer.
\newblock {\em Journal of Mathematical Biology}, 15:267--273, 1982.

\bibitem[PB98]{PoBl98}
J.~Pollack and A.~Blair.
\newblock Co-evolution in the successful learning of backgammon strategy.
\newblock {\em Machine Learning}, 32:225–24, 1998.

\bibitem[Ros62]{Rose62}
F.~Rosenblatt.
\newblock {\em Principles of Neurodynamics}.
\newblock Spartan Books, New York, 1962.

\bibitem[San89]{Sanger89}
T.~D. Sanger.
\newblock Optimal unsupervised learning in a single-layered linear feedforward
  network.
\newblock {\em Neural Networks}, 2:459--473, 1989.

\bibitem[SJK09]{Szub09}
Marcin Szubert, Wojciech Jaskowski, and Krzysztof Krawiec.
\newblock Coevolutionary temporal difference learning for othello.
\newblock In {\em Proceedings of the 5th international conference on
  Computational Intelligence and Games}, CIG'09, pages 104--111, Piscataway,
  NJ, USA, 2009. IEEE Press.

\bibitem[SSM98]{Scho98}
Bernhard Sch\"{o}lkopf, Alexander Smola, and Klaus-Robert M\"{u}ller.
\newblock Nonlinear component analysis as a kernel eigenvalue problem.
\newblock {\em Neural Comput.}, 10:1299--1319, July 1998.

\bibitem[Tes92]{Tesa92}
Gerald Tesauro.
\newblock Practical issues in temporal difference learning.
\newblock {\em Mach. Learning}, 8:257--277, 1992.

\bibitem[Tes94]{Tesa94a}
Gerald Tesauro.
\newblock {TD}-gammon, a self-teaching backgammon program, achieves
  master-level play.
\newblock {\em Neural Computation}, 6:215--219, 1994.

\bibitem[vEvW05]{vEck05}
N.J. van Eck and M.~van Wezel.
\newblock Reinforcement learning and its application to othello.
\newblock Technical Report EI 2005-47, Econometric Institute Report, Erasmus
  University Rotterdam, 2005.

\bibitem[Wis98]{Wis98a}
L.~Wiskott.
\newblock
  \href{http://itb.biologie.hu-berlin.de/~wiskott/Publications/Wis98a-InvarianceManifolds-JSNC.ps.gz}{Learning
  Invariance Manifolds}.
\newblock In {\em Proc.\ of the 5th Joint Symp.\ on Neural Computation, May 16,
  San Diego, CA}, volume~8, pages 196--203, San Diego, CA, 1998. Univ.\ of
  California.

\bibitem[WS02]{WisSej2002}
L.~Wiskott and T.~Sejnowski.
\newblock
  \href{http://itb.biologie.hu-berlin.de/~wiskott/Publications/WisSej2002-LearningInvariances-NC.ps.gz}{Slow
  Feature Analysis: Unsupervised Learning of Invariances.}
\newblock {\em Neural Computation}, 14(4):715--770, 2002.

\end{thebibliography}

\newcommand{\etalchar}[1]{$^{#1}$}

\end{document}